\documentclass[comment, prl,twocolumn,nofootinbib, preprintnumbers, superscriptaddress]{revtex4-2}

\usepackage{booktabs}
\usepackage{multirow}
\usepackage{amsmath,amssymb,slashed,braket,bm}
\usepackage{graphicx}
\usepackage{cancel}
\usepackage{epstopdf}
\usepackage{float}
\usepackage[colorlinks=true,
linkcolor=blue,
urlcolor=blue,
citecolor=green,          
bookmarks=true,
bookmarksnumbered=true,
breaklinks=true,
pdfpagemode=Fullscreen,
pdfstartview=FitBH]{hyperref}

\usepackage[normalem]{ulem}
\usepackage{color}

\definecolor{Orange}{cmyk}{0,0.61,0.87,0}
\definecolor{JungleGreen}{cmyk}{0.99,0,0.52,0}
\definecolor{OliveGreen}{cmyk}{0.64,0,0.95,0.40}
\definecolor{Brown}{cmyk}{0,0.81,1,0.60}
\definecolor{RoyalBlue}{cmyk}{0.71,0.53,0,0.12}
\definecolor{Gray}{cmyk}{0,0,0,0.40}
\definecolor{LightPink}{cmyk}{0.0,0.25,0,0}
\definecolor{LLightPink}{cmyk}{0.0,0.10,0,0}
\definecolor{LightBlue}{cmyk}{0.25,0,0,0}
\definecolor{LightGray}{cmyk}{0,0,0,0.2}

{}
{}
{}

{}

\usepackage{xcolor}
\definecolor{gesfpurple}{rgb}{0.47,0.19,0.42}

\definecolor{gesflanse}{rgb}{0.00,0.50,0.50}

\definecolor{gesfblue}{rgb}{0.08,0.42,0.76}

\definecolor{gesfred}{rgb}{1,0,0}

\definecolor{gesfwhite}{rgb}{1,1,1}

\definecolor{gesfblack}{rgb}{0,0,0}

\newcommand{\geqn}[1]{Eq.\,\hypersetup{linkcolor=blue}(\ref{#1})\hypersetup{linkcolor=blue}}
\newcommand{\gfig}[1]{{\hypersetup{linkcolor=violet}Fig.\,\ref{#1}\hypersetup{linkcolor=blue}}}

\newcommand{\MP}{m_\phi}
\newcommand{\ml}{m_{\rm lightest}}

\begin{document}

\title{Lower Bound on the Lightest Neutrino Mass
with Resonant Neutrino Self-Interactions
}

\author{Shao-Feng Ge}
\email{gesf@sjtu.edu.cn}
\affiliation{State Key Laboratory of Dark Matter Physics, Tsung-Dao Lee Institute \& School of Physics and Astronomy, Shanghai Jiao Tong University, China}
\affiliation{Key Laboratory for Particle Astrophysics and Cosmology (MOE) 
\& Shanghai Key Laboratory for Particle Physics and Cosmology, Shanghai Jiao Tong University, Shanghai 200240, China}

\author{João Paulo Pinheiro}
\email{joaopaulo.pinheiro@sjtu.edu.cn}
\affiliation{State Key Laboratory of Dark Matter Physics, Tsung-Dao Lee Institute \& School of Physics and Astronomy, Shanghai Jiao Tong University, China}
\affiliation{Key Laboratory for Particle Astrophysics and Cosmology (MOE) 
\& Shanghai Key Laboratory for Particle Physics and Cosmology, Shanghai Jiao Tong University, Shanghai 200240, China}

\author{Shaoyang Qin}
\email{shaoyang.qin@sjtu.edu.cn}
\affiliation{State Key Laboratory of Dark Matter Physics, Tsung-Dao Lee Institute \& School of Physics and Astronomy, Shanghai Jiao Tong University, China}
\affiliation{Key Laboratory for Particle Astrophysics and Cosmology (MOE) 
\& Shanghai Key Laboratory for Particle Physics and Cosmology, Shanghai Jiao Tong University, Shanghai 200240, China}

\author{Ruitao Sun}
\email{rt\_sun@sjtu.edu.cn}
\affiliation{State Key Laboratory of Dark Matter Physics, Tsung-Dao Lee Institute \& School of Physics and Astronomy, Shanghai Jiao Tong University, China}
\affiliation{Key Laboratory for Particle Astrophysics and Cosmology (MOE) 
\& Shanghai Key Laboratory for Particle Physics and Cosmology, Shanghai Jiao Tong University, Shanghai 200240, China}

\author{Oleg Titov}
\email{titov\_o@sjtu.edu.cn}
\affiliation{State Key Laboratory of Dark Matter Physics, Tsung-Dao Lee Institute \& School of Physics and Astronomy, Shanghai Jiao Tong University, China}
\affiliation{Key Laboratory for Particle Astrophysics and Cosmology (MOE) 
\& Shanghai Key Laboratory for Particle Physics and Cosmology, Shanghai Jiao Tong University, Shanghai 200240, China}

\begin{abstract}
We show the possibility of using the resonant
$s$-channel scattering of ultra high energy
astrophysical neutrinos with the low energy
cosmic neutrino background to put a lower bound
on the lightest neutrino mass. This can be realized and illustrated
by neutrino self-interactions with a light scalar mediator of $\mathcal O(1)$\,MeV
mass.
After analyzing the kinematics and cross section
features of the $s$-channel resonance, we derive
from first principles the double transport equations
with double regeneration terms for the coupled 
evolution of neutrinos and anti-neutrinos.
The simulated event spectrum is then compared
with the HESE 7.5 year data observed at IceCube
to give a clear lower bound on the lightest
neutrino mass.

\end{abstract}

\maketitle

\textbf{Introduction} -- 
The massive neutrinos are the first new physics
beyond the Standard Model (SM) of particle
physics \cite{ParticleDataGroup:2026mpi}. Various evidence have been
established by the neutrino oscillation experiments
with atmospheric \cite{Super-Kamiokande:1998kpq, Super-Kamiokande:2023ahc, IceCubeCollaboration:2023wtb},
solar \cite{ GALLEX:1998kcz, SNO:2002tuh,Super-Kamiokande:2005wtt,Super-Kamiokande:2008ecj,Borexino:2008fkj,Super-Kamiokande:2010tar,SNO:2011hxd,Borexino:2017rsf,BOREXINO:2020aww}, accelerator \cite{K2K:2002icj,MINOS:2013utc,T2K:2023smv,T2K:2025yoy},
and reactor \cite{KamLAND:2002uet, KamLAND:2010fvi, DoubleChooz:2011ymz, DayaBay:2012fng, RENO:2012mkc, KamLAND:2013rgu, DayaBay:2022orm,
JUNO:2025gmd,JUNO:2026tap} neutrinos.
Especially, the associated leptonic CP phases
are of profound importance
for explaining the baryon asymmetry observed in our
Universe through the leptogenesis mechanism
\cite{Fukugita:1986hr, Davidson:2008bu}. The tiny neutrino masses
are important keys for fully understanding
the evolution of our Universe and the very
existence of our matter world.
However, we still do not know the absolute scale
of the neutrino masses $m_i$ with $i = 1$, 2, 3.
The neutrino oscillation experiments can measure
the solar mass squared difference
$\Delta m^2_s \equiv m^2_2 - m^2_1 = 7.388{\pm 0.078} \times 10^{-5}$\,eV$^2$ \cite{JUNO:2026tap}
and its atmospheric counterpart
$|\Delta m^2_a| \equiv |m^2_3 - m^2_1| = 2.521^{+0.026}_{-0.018} \times 10^{-3}$\,eV$^2$ 
at the percentage level \cite{Esteban:2024eli,Esteban:2026phq}
with the absolute scale appears only at higher orders
\cite{Alves:2026nce}.
Without knowing the sign of $\Delta m^2_a$, the
neutrino masses can follow the normal ordering (NO)
with $m_1 < m_2 < m_3$ or the inverted ordering (IO)
with $m_3 < m_1 < m_2$. The lightest mass ($m_1$ or
$m_3$) could be massless.

The $\beta$ decay end-point spectrum depends on the
effective mass $m_\beta \equiv \sqrt{\sum_i |U_{ei}|^2 m^2_i}$
\cite{Giunti:2007ry} where $U$ is the neutrino mixing matrix.
Measuring the beta decay spectrum can then probe the
neutrino mass. While KATRIN has used the tritium beta decay
to put an upper bound $m_\beta < 0.45$\,eV at 90\%\,C.L.
\cite{KATRIN:2019yun,KATRIN:2021uub,KATRIN:2024cdt},
Project~8 aims at a sensitivity of about 40\,meV
\cite{Project8,Project8Roadmap}. If neutrinos are Majorana
particles, the neutrinoless double beta decay
measures the effective Majorana mass $m_{\beta\beta} \equiv \sum_i U^2_{ei} m_i$ 
\cite{Bilenky:1987ty,Dolinski:2019nrj,Agostini:2022zub}
and the current searches find
$\langle m_{\beta\beta}\rangle < (28$--$122)$\,meV
\cite{KamLAND-Zen:2024eml,PandaX:2024fed,CUORE:2024ikf,AMoRE:2024loj,LEGEND:2025jwu,PandaX:2025yly}.
Although vanishing $m_{\beta\beta}$ can also constrain the
absolute mass scale \cite{Lindner:2005kr,Ge:2016tfx,Xing:2016ymd,Penedo:2018kpc,Cao:2019hli,Denton:2023hkx,Ge:2025cky}
and the two Majorana CP phases
\cite{Xing:2003jf,Ge:2016tfx,Xing:2016ymd,Ge:2019ldu,Ge:2025cky},
it has not been experimentally reached yet 
\cite{Agostini:2017iyd,nEXO:2017nam,EXO-200:2019rkq,DARWIN:2020jme,Lee:2020rjh,NEXT:2020amj,CUPID:2020aow,GERDA:2020xhi,CUORE:2021mvw,SNO:2021xpa,nEXO:2021ujk,LEGEND:2021bnm,KamLAND-Zen:2022tow,CUPID:2022wpt,Majorana:2022udl,KamLAND-Zen:2024eml,PandaX:2024fed,CUORE:2024ikf,AMoRE:2024loj,LEGEND:2025jwu,PandaX:2025yly}.

The current most stringent upper bound comes from cosmology.
Although the neutrino mass is tiny, the large amount of the
cosmic relic neutrinos can contribute sizable energy and
leave observable effects on the cosmic microwave background
(CMB) and large scale structures
\cite{Lesgourgues:2006nd,Lesgourgues:2013sjj,TopicalConvenersKNAbazajianJECarlstromATLee:2013bxd,Dvorkin:2019jgs,DiValentino:2024xsv}.
The combination of CMB, DESI baryon acoustic oscillation,
and the oscillation data gives 
$\ml < 0.023$\,eV (NO) and
$\ml < 0.024$\,eV (IO)
within the $\Lambda$CDM model
\cite{Planck:2018vyg,DESI:2024mwx,Elbers:2025vlz,DESI:2025zgx}.
The gravitational focusing of the cosmic neutrino fluid
by dark matter halos can provide a third independent
cosmological measurement with a different mass dependence
$\sum m_i^4$ \cite{Ge:2023nnh,Ge:2024kac}.

There are another two approaches that can also in
principle measure the neutrino absolute masses.
By capturing the cosmic neutrino background
(C$\nu$B) neutrinos on tritium,
$\nu_e + {}^3{\rm H} \to {}^3{\rm He} + e^-$, the final-state
electron spectrum can extend beyond the beta decay spectrum
end point by $2 m_i$ as a line structure
\cite{Cocco:2007za,Long:2014zva,PTOLEMY:2018jst}. The second
is the radiative emission of neutrino pairs from atomic
transitions whose photon spectrum has various turning
points at $(m_i+m_j)^2$ \cite{Fukumi:2012rn,Dinh:2012qb}.

However, the current measurements can only put an
upper bound on the neutrino mass scale or equivalently
the lightest mass. 
Establishing a practical way of bounding the
lightest mass ($m_1$ for NO or $m_3$ for IO)
would complete the constraints from both ends.
We point out that the kinematics
of resonant scattering between the C$\nu$B and
the ultra-high-energy (UHE) neutrinos through
a light mediator can provide such an observable.

\vspace{2mm}
\textbf{Kinematics of Resonant Scattering}
--
For the scattering of an UHE neutrino of energy $E_\nu$
with a C$\nu$B neutrino $\nu_i$ with momentum $\bm p$ and energy
$E_i \equiv \sqrt{\bm p^2 + m_i^2}$, the squared
center-of-mass energy,
\begin{equation}
s
\simeq
2 E_\nu \left( E_i - |\bm p| \cos\theta \right),
\label{eq:s_general}
\end{equation}
where $\theta$ is the scattering angle. If the two
neutrinos interact through a mediator $\phi$
with mass $\MP$ in the $s$ channel, the cross section
is resonantly enhanced at $s = \MP^2$.

For a non-relativistic relic neutrino
$|\bm p| \ll E_i \approx m_i$, the resonance
happens at a single energy $E_\nu \approx m^2_\phi / 2 m_i$.
This can be naturally realized for at least two
massive neutrinos ($m_2$ and $m_3$ for NO or
$m_1$ and $m_2$ for IO). Even for the smaller splitting,
$\sqrt{\Delta m^2_s} = 8.68 \times 10^{-3}$\,eV
is much larger than the current C$\nu$B temperature
at $T_\nu = 1.9\,\mbox{K}$ ($1.68 \times 10^{-4}$\,eV)
\cite{Lesgourgues:2013sjj,ParticleDataGroup:2024cfk}.

On the other hand, the lightest neutrino could
still be relativistic if its mass is around or
even smaller than the C$\nu$B temperature.
The resonance would happen in a much broader
energy range $E^-_\nu \leq E_\nu \leq E^+_\nu$,
\begin{equation}
E^\pm_\nu
\equiv
\frac{\MP^2}{2 (E_i \mp |\bm p|)}
=
\frac {\MP^2}{2 m^2_i} (E_i \pm |\bm p|),
\label{eq:res_window}
\end{equation}
Since the product $\sigma v_{\rm rel}$ of cross
section $\sigma$ and relative velocity $v_{\rm rel}$
is only a function of $s$ which is fixed by the mediator
mass at $m^2_\phi$, there is no actual dependence on
the initial-state $\cos \theta$ in \geqn{eq:s_general}.
Then the UHE energy $E_\nu$ would have
a flat distribution between $E^-_\nu$
and $E^+_\nu$.

For an ultra-relativistic relic neutrino with
$E_i \approx |\bm p|$, the lower and upper limits
can be approximated as,
\begin{align}
E^-_\nu
\approx
\frac {m^2_\phi}{4 |\bm p|},
\qquad
E^+_\nu
\approx
\frac {m^2_\phi}{m^2_i} |\bm p|.
\label{eq:edges_obs}
\end{align}
While the lower limit $E^-_\nu$
is fixed, the upper one can vary with the lightest
neutrino mass $m_i$.
The ratio $E^+_\nu / E^-_\nu \approx 4 |\bm p|^2 / m^2_i$
between the upper and lower limits is inversely
proportional to $m^2_i$ and is independent of the
mediator mass $m_\phi$.

In between, the transition case with
$|\bm p| \sim m_i$ should have a flat region
whose width decreases with the
lightest neutrino mass. Using the exact
formula in \geqn{eq:res_window},
\begin{align}
  \frac {E^+_\nu}{E^-_\nu}
=
  \frac {E_i + |\bm p|}{E_i - |\bm p|},
\quad \mbox{or} \quad
  m_i
=
  \frac {2 \sqrt{E^-_\nu E^+_\nu}}{E^+_\nu - E^-_\nu}
  |\bm p|.
\label{eq:Elimits-ratio}
\end{align}
For a non-relativistic relic neutrino,
the dip width $E^+_\nu - E^-_\nu$ should vanish with
its momentum $\bm p$ to maintain a finite mass $m_i$.
With resonance, the incoming UHE neutrino is
absorbed to regenerate neutrinos at much lower
energy, leaving behind a wide dip in the UHE
neutrino flux. Observing the dip boundaries
$E^-_\nu$ and $E^+_\nu$, one may reconstruct
the lightest neutrino mass.

Note that the cosmic relic neutrino momentum
follows the Fermi-Dirac distribution
$1 / (e^{|\bm p| / T_\nu} + 1)$ with $T_\nu = 1.9$\,K
nowadays. Although not fixed, its average
$\langle |\bm p| \rangle \simeq 3.15\,T_\nu \simeq 5.3\times10^{-4}$\,eV,
can be used to infer the lightest neutrino
mass statistically. 

Both the cosmic relic and UHE neutrinos have
traveled cosmological distances. It is then
necessary to take the effect of cosmological
expansion into account. The relic momentum
$\bm p$ used earlier should be replaced by
$|\bm p| = |\bm p_0|(1+z)$ while the UHE
energy by $E_\nu = E_0 (1+z)$ with the subscript
``0'' denoting the values in the current Universe.
For sources tracing the star formation history
with redshift $z\sim 1$--$2$ \cite{Yuksel:2008cu},
the relic momentum is $|\bm p| \sim 10^{-3}$\,eV
which is still much smaller than $\sqrt{\Delta m^2_s}$
and the basic features pointed out earlier
still apply.

\vspace{2mm}
\textbf{Realization with Neutrino Self-Interactions} --
The mechanism of using the UHE neutrino spectrum for
probing the lightest neutrino mass requires a neutral
mediator. The minimal choice is a light scalar coupled
only with neutrinos. We consider a real scalar $\phi$ coupled
with Majorana active neutrinos \cite{Berryman:2022hds},
\begin{equation}
\mathcal{L}
\supset
- \frac{1}{2} m_\phi^2 \phi^2
- \frac{1}{2}\, g_{\alpha\beta}\, \phi\,
\overline{\nu^c_{L\alpha}}\, \nu_{L\beta} + {\rm h.c.},
\label{eq:Lagrangian}
\end{equation}
where $g_{\alpha\beta} = g_{\beta\alpha}$ is symmetric
for Majorana neutrinos. We adopt the minimal flavor structure 
$g_{\alpha\beta} = g_{\tau\tau}\,\delta_{\alpha\tau}\delta_{\beta\tau}$
whose laboratory bounds are the weakest
\cite{Blinov:2019gcj,Brdar:2020nbj,Esteban:2021tub,Berryman:2022hds}.
Such neutrino self-interactions ($\nu$SI)
leave the charged leptons, quarks and the electroweak sector untouched
which can be realized by seesaw-inspired
UV completions
\cite{Berryman:2018ogk,deGouvea:2019qaz,Dev:2021axj,Berryman:2022hds}.
Since both the cosmic relic and astrophysical UHE
neutrinos have been existing for quite long time,
they should have already lost coherence and need
to be treated as mass eigenstates,
$ \mathcal L
\supset
- \frac{1}{2}\, \phi\, \overline{\nu_i}
  \left( g_{ij} P_L + g^*_{ij} P_R \right) \nu_j$
with
$\nu_{L \alpha} \equiv U_{\alpha i}\,\nu_{Li}$
and couplings
$g_{ij} \equiv g_{\tau\tau} U_{\tau i} U_{\tau j}$.

The neutrino scattering can happen as
$\nu \bar \nu \rightarrow \phi \rightarrow \nu \bar \nu$
through the $s$ channel with cross section \cite{Esteban:2021tub},
\begin{equation}
\sigma_{ij\to kl}(s)
= \frac{|g_{ij}|^2 |g_{kl}|^2}{16\pi}\,
\frac{s}{(s - m_\phi^2)^2 + m_\phi^2 \Gamma_\phi^2},
\label{eq:sigma_BW}
\end{equation}
where
$\Gamma_\phi \equiv \frac{m_\phi}{16\pi} \sum_{k,l} |g_{kl}|^2$
is the decay rate of $\phi$. It is also possible to have the
$t$ and $u$ channels. However, the $s$ channel can be
significantly enhanced in the resonance region to
dominate over the other channels.

For an MeV-scale mediator, the resonance with the
C$\nu$B falls within $10^{5} \sim 10^{7}$\,GeV
\cite{Ng:2014pca,Ioka:2014kca,Ibe:2014pja,Blum:2014ewa,Kelly:2018tyg}
which is covered by the IceCube High-Energy Starting Event (HESE)
sample \cite{IceCube:2020wum}.
The HESE spectrum does not follow exactly the single power law expected
from astrophysical acceleration\,\cite{Fermi:1949ee,Gaisser:2016uoy,Hill:2025vmu}
and its features have been interpreted as hints of
$\nu$SI\,\cite{Kamada:2015era,Bustamante:2020mep,Mazumdar:2020ibx,Carpio:2021jhu,Esteban:2021tub,delaVega:2024pbk,Francener:2025apz}.
In these analyses, as well as in the forecasts for
GRAND\,\cite{Machado:2025ltu,Leal:2025eou}, the search
target is the mediator. The kinematics
is then reduced to the two limits of \geqn{eq:res_window}:
a relic at rest, $\bm p = 0$
\cite{Esteban:2021tub,Bhattacharya:2026dng},
or a massless one, $|\bm p| = E_i$ \cite{Wang:2025qap}.
However, the transition region has not been
fully explored yet.

\vspace{2mm}
\textbf{Resonant Transport via $\nu$SI} --
The astrophysical UHE neutrinos are produced from
violent astrophysical processes such as AGN
\cite{Stecker:1991vm,Murase:2022feu}. Without
measuring the baseline, a neutrino telescope can see
all the neutrino sources along the line of sight\,\cite{Esteban:2021tub, IceCube:2020wum}, 
\begin{align}
  \mathcal S^{\nu/\bar\nu}_i(E, z)
\equiv
  F(z)\, f_{\nu/\bar\nu}
  \sum_\alpha |U_{\alpha i}|^2 f_\alpha\,
  \Phi_0 \left(\frac{E}{E_{\rm ref}}\right)^{-\gamma},
\label{eq:source}
\end{align}
where $E$ is the neutrino energy at redshift $z$.
With a reference energy $E_{\rm ref} = 100$\,TeV, the
source spectrum follows power law with index $\gamma$.
We take the source flavor composition
$f_\alpha$ as $(f_e, f_\mu, f_\tau) = (1/3, 2/3, 0)$
from pion decay \cite{Esteban:2026vzh,Beacom:2003nh,Pakvasa:2007dc}. The
neutrino and antineutrino fractions are
$(f_\nu, f_{\bar\nu}) = (1/2, 1/2)$ for $pp$ sources and
$(2/3, 1/3)$ for $p\gamma$ sources\,\cite{Skrzypek:2025tmg,Liu:2023lxz}.
Losing coherence due to long distance,
the source spectrum should then be
transformed with the mixing matrix
elements $U_{\alpha i}$ and the decoherence is
reflected in taking its magnitude as $|U_{\alpha i}|^2$. 

The redshift distribution $F(z)$
should follow the star formation rate 
$F_{\rm SFR}$ \cite{Yuksel:2008cu},
\begin{align}
\hspace{-3mm}
F(z)
\equiv
F_0
\left[
(1 + z)^{a \eta}
+ \left( \frac {1+z} B \right)^{b \eta}
+ \left( \frac {1+z} C \right)^{c \eta}
\right]^{\frac 1 \eta} \hspace{-2mm},
\end{align}
where $a = 3.4$, $b = -0.3$, and $c = -3.5$
are the logarithmic slopes with $\eta \approx - 10$
to smooth the transitions. The breaks at
$z_1 = 1$ and $z_2 = 4$ take $B = (1+z_1)^{1 - \frac a b} \approx 5000$
and $C = (1 + z_1)^{\frac {b-a} c} (1 + z_2)^{1 - \frac b c} \approx 9$.
However, the normalization $F_0$ for the
UHE neutrino flux cannot be directly
fixed and needs to fit the IceCube data
together with other model parameters.

When propagating through space, the UHE neutrinos
are diluted by the Hubble expansion
and experience scattering with C$\nu$B.
With both neutrinos and anti-neutrinos,
the comoving diffused fluxes
$\phi^{\nu / \bar \nu}_i \equiv a^3 v d n^{\nu / \bar \nu}_i / d |\bm p|$
evolve with two transport equations,
\begin{subequations}
\begin{align}
  \partial_z \phi_i^\nu
& =
  \frac {- \mathcal S_i^\nu
         - H \partial_E (E \phi_i^\nu)
         + \Gamma^\nu_i \phi^\nu_i 
         - \Gamma^{\nu \nu}_i
         - \Gamma^{\bar \nu \nu}_i}
        {H (1+z)},
\label{eq:transport_nu} 
\\
  \partial_z \phi_i^{\bar \nu}
& =
  \frac {- \mathcal S_i^{\bar \nu}
         - H \partial_E (E \phi_i^{\bar \nu})
         + \Gamma^{\bar \nu}_i \phi^{\bar \nu}_i 
         - \Gamma^{\bar \nu \bar \nu}_i 
         - \Gamma^{\nu \bar \nu}_i}
        {H (1 + z)}.
\label{eq:transport_antinu} 
\end{align}
\label{eq:transport} 
\end{subequations}
Although sharing
the same expression, the comoving diffused flux
$\phi^{\nu / \bar \nu}_i = a^3 v \frac {|\bm p|^2}{2 \pi^2} f^{\nu / \bar \nu}_i(|\bm p|)$
is different from the usually adopted comoving number density per
unit energy (which is actually comoving spectrum)
\cite{Ng:2014pca,Esteban:2026vzh}. The concept
of diffused comoving flux is more appropriate
for evaluating the observed event rate.
For ultra-relativistic UHE neutrinos,
one may simply take $v \approx 1$ and
$|\bm p| \approx E$.

The cosmic expansion is reflected by the
Hubble rate $H$, in particular the second
term of \geqn{eq:transport}. If only
the source term $\mathcal S$ and Hubble expansion
are considered, the UHE flux can be solved
analytically \cite{Ng:2014pca, Esteban:2026vzh},
\begin{equation}
\hspace{-2mm}
  \phi^{\nu / \bar\nu}_i(z, E_0)
= 
  \frac 1 {1 + z}
  \int^{z_{\rm max}}_z
  \frac {\mathcal S^{\nu/\bar\nu}_i[z', E_0 (1+z')]}{H(z')} d z',
\end{equation}
where $E_0$ is the neutrino energy today
when arriving at the Earth. The integration
from the redshift upper limit $z_{\rm max} = 2$
to some point $z$ clearly indicates the
initial condition $\phi^{\nu/\bar\nu}_i(z_{\rm max}, E_0) = 0$
and the accumulative contribution from
various sources along the path.

During propagation, the UHE neutrinos scatter
with the non-relativistic C$\nu$B.
In particular, the $s$-channel scattering
$\nu_{\rm UHE}(p) + \bar \nu_{\rm C \nu B}(k) \rightarrow \nu(p') + \bar \nu(k')$
can induce significant resonance as summarized
in \geqn{eq:sigma_BW}. While absorbing one initial
UHE neutrino from its energy and flavor, one neutrino
and one anti-neutrino are regenerated with
lower energy. The absorption and regeneration
are described by the corresponding rates $\Gamma^a_i$ and
$\Gamma^{ab}_i$ ($a, b = \{\nu , \bar \nu\}$), 
respectively, 
\begin{subequations}
\begin{align}
  \Gamma^a_i
& \equiv \hspace{-1mm}
  \sum_{jkl} \hspace{-1mm}
  \int f^j_{\rm C \nu B}(\bm k)
  \sigma^a_{ij \to kl} \frac {d^3 \bm k}{(2 \pi)^3},
\label{eq:Gamma-}
\\
  \Gamma^{ab}_i
& \equiv \hspace{-1mm}
  \sum_{klj} \hspace{-1mm}
  \int \phi^a_k(|\bm p'|) d |\bm p'|
  \hspace{-1mm}
  \int \frac {f^l_{\rm C \nu B}(\bm k') d^3 \bm k'}{(2 \pi)^3}
  \frac {d \sigma^{a \rightarrow b}_{kl \to ij}}{d |\bm p|}.
\label{eq:Gamma+}
\end{align}
\end{subequations}
\begin{figure}
\centering
\includegraphics[width=0.9\linewidth]{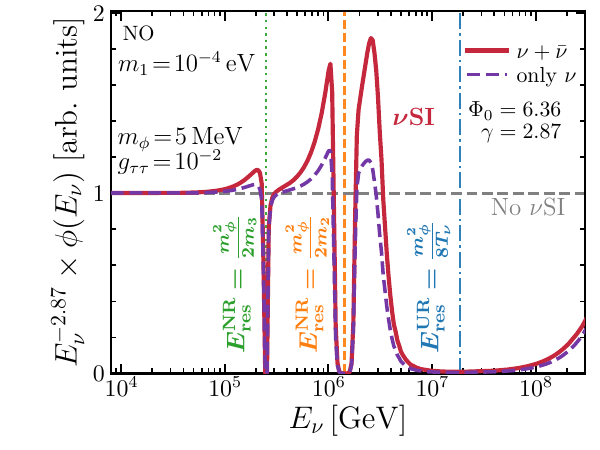}
\caption{The total neutrino flux arriving at Earth 
for the normal ordering (NO). 
The red solid and purple dashed curves denote the results
with the double regeneration terms
($\nu + \bar \nu$) or a single one
(only $\nu$).
For comparison, the horizontal grey dashed line shows
the case without $\nu$SI. 
The colored vertical lines mark the resonance energies
$E_{\rm res}$ for the relativistic $m_1$ (blue dash-dotted)
as well as the non-relativistic $m_2$ (orange dashed)
and $m_3$ (green dotted). 
}
\label{fig:Spectrum}
\end{figure}
Taking a single $\Gamma^{ab}$ in the transport equation
\cite{Esteban:2021tub} underestimates
the flux arriving at Earth as shown 
in \gfig{fig:Spectrum}. For a relativistic $m_1 = 10^{-4}$\,eV,
the other two mass eigenvalues are non-relativistic
$m_2 \simeq 8.68 \times 10^{-3}$\,eV and
$m_3 \simeq 5.01 \times 10^{-2}$\,eV with
the best-fit values of $\Delta m^2_s$ and $\Delta m^2_a$ quoted
above \cite{Esteban:2024eli,NuFIT6.1}.
For non-relativistic $\nu_2$ and $\nu_3$,
the resonance happens at the middle point
$E_{\rm res} \equiv (E^+_\nu + E^-_\nu)/2 = m^2_\phi E_i / 2 m^2_i$
with values $E^{\rm NR}_{2,3} \simeq \{1.44\times10^{6},\ 2.50\times10^{5}\}$\,GeV.
However, the bottom of the wide resonance dip
for relativistic neutrinos happens at
$m^2_\phi / 8 T_\nu$. Using the log
scale, the nominal difference between the dip edges
is essentially the logarithm of the ratio
$E^+_\nu / E^-_\nu = (1 + v_i) / (1 - v_i)$
according to \geqn{eq:Elimits-ratio}. For a typical
C$\nu$B momentum $|\bm p| = \langle p \rangle$ today,
these masses give the relic neutrino velocities
$v_{1,2,3} \simeq \{0.98,\ 0.06,\ 0.01\}$. For
the non-relativistic $\nu_2$ and $\nu_3$,
$\log_{10} (E^+_\nu / E^-_\nu) \approx 2 v_i \simeq \{0.12,\ 0.02\}$
explaining why the $\nu_2$ dip is visibly
broader than the $\nu_3$ one. For the relativistic
$\nu_1$, its $\log_{10} (E^+_\nu / E^-_\nu) \approx 2$
indicating that the dip spans more than one order.
The natural width of the resonance,
$\Gamma_\phi/\MP = g_{\tau\tau}^2/16\pi \sim 10^{-6}$,
is negligible.

Another prominent feature is
that each dip is accompanied by a sharp peak at its
low-energy edge due to the regeneration effect.
Adopting the double regeneration terms
($\nu + \bar \nu$) in
\geqn{eq:transport} enhances these bumps.

\begin{figure}
\centering
\includegraphics[width=0.9\linewidth]{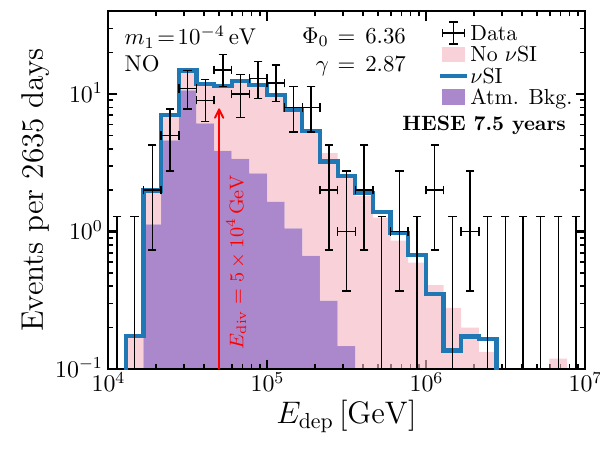}
\caption{Fitting the HESE 7.5 years data (black points with error bar)
as a function  of the deposit energy
$E_{\rm dep}$ for
the standard scenario without $\nu$SI (pink shaded)
and the $\nu$SI prediction (blue solid line).
The purple shaded histogram shows the total
atmospheric background, neutrinos and muons.}
\label{fig:Event_Number}
\end{figure}

\vspace{2mm}
\textbf{Lower Bound on the Lightest Neutrino Mass} --
We compare the predicted flux
$\phi_\alpha(E) \equiv \phi_\alpha(z=0,E)$ at Earth
with the 102 events collected over $T=2635$ days for
the public HESE 7.5-year release\,\cite{IceCube:2020wum}.
The data in \gfig{fig:Event_Number} has
a deficit at the deposit energy
$E_{\rm dep} \sim 5 \times 10^5$\,GeV, an excess near $10^6$\,GeV, a
second deficit around $2 \times 10^6$\,GeV, and no events
above \cite{IceCube:2020wum,Esteban:2021tub}.
The narrow $\nu_{2,3}$ deficits, each
followed by a recovery of the flux, can
reproduce this alternating pattern.
In addition, the $\nu_1$ suppression removes
events continuously from $3\times10^{6}$\,GeV
upward.

The expected events are added to the atmospheric neutrino
and muon backgrounds \cite{IceCube:2020wum}.
For the benchmark in \gfig{fig:Spectrum}, the
$\nu_{2,3}$ resonances at $E^{\rm NR}_3$ and $E^{\rm NR}_2$
leave narrow deficits around $(2\text{--}3)\times10^{5}$\,GeV
and $(1\text{--}2)\times10^{6}$\,GeV after smearing
when mapping from the neutrino true energy
$E$ to the deposit energy $E_{\rm dep}$. Consequently,
the narrow $\nu_{2,3}$ windows are partially filled and
the regeneration bumps just below them are largely washed out.
Only the two-decade $\nu_1$ window can instead survive and
suppress the high-energy tail above around
$3 \times 10^{6}$\,GeV. This corresponds to the lower edge
$\MP^2/4\langle p_0 \rangle (1+z)^2 \simeq 3\times10^{6}$\,GeV
of \geqn{eq:edges_obs} for sources at $z \sim 1$.

Until now, the UHE neutrinos have only been measured by
IceCube. Both the flux normalization
$\Phi_0$ and the power index $\gamma$ need to be
determined by fitting the IceCube data. For all cases
in \gfig{fig:Event_Number}, the
astrophysical flux is fixed at the 
HESE best fit without $\nu$SI, $\Phi_0=6.36$
and $\gamma=2.87$\,\cite{IceCube:2020wum}
to make illustration. 

\begin{figure}
\centering
\includegraphics[width=\linewidth]{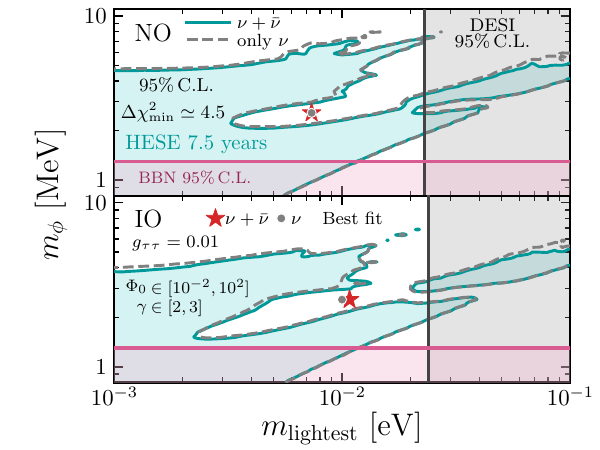}
\caption{The excluded regions (shaded) on the $(\ml, \MP)$ plane
at 95\%\,C.L. by the HESE 7.5-year data considering
the neutrino and anti-neutrino double (cyan solid) or
neutrino only (gray dashed) transport equations
with normal (upper) or inverted (lower) ordering.
The red star and gray ball mark the corresponing best-fit points.
For comparison, the black solid line shows the 95\% upper
bound from DESI DR2 BAO combined with CMB assuming $\Lambda$CDM
and oscillation data \cite{Elbers:2025vlz}.
The $\Delta \chi^2_{\rm min} \equiv 
 \chi^2_{\rm std,min}
-
\chi^2_{\nu\mathrm{SI,min}}$
indicates the neutrino self-interaction
improvement than the standard scenario.
}
\label{fig:Constraint}
\end{figure}

We confront the predicted signal distributions with
the HESE data using the likelihood of the public release
and the systematic uncertainties \cite{IceCube:2020wum}.
Besides the lightest neutrino mass $m_{\rm lightest}$
and the mediator mass $m_\phi$, we marginalize the
normalization $\Phi_0 \in [10^{-2},10^2]$ and the
spectral index $\gamma \in [2,3]$ with fixed
$g_{\tau\tau} = 0.01$.
The very light neutrino mass range can indeed be
excluded and the lightest mass receives a
lower bound as shown in \gfig{fig:Constraint}.

The boundaries follow from the resonance kinematics.
To the lower right, the excluded region ends along
$\MP^2 \simeq 2 E_{\rm div}\,\ml$ with the
lightest-state dip $E^{\rm NR}_{\rm l} \approx \MP^2/2\ml$
around the division energy $E_{\rm div} = 5 \times 10^4$\,GeV
below which the atmospheric neutrino
background dominates as shown in \gfig{fig:Event_Number}.
Beyond this line, the resonance falls into the background
dominating region and the data lose sensitivity. For
$\ml \lesssim 3\times10^{-3}$\,eV, the upper boundary
becomes flat at $\MP \simeq 4.7$\,MeV (NO) and
$\simeq 4$\,MeV (IO). This is close to the
ultra-relativistic
regime of \geqn{eq:edges_obs} where the predicted
spectrum no longer depends on $\ml$.

The allowed band between these boundaries contains the
best fit at $\ml = 7.4\times10^{-3}$\,eV (NO) and
$1.1\times10^{-2}$\,eV (IO) with $\MP \simeq 2.6$\,MeV
in both orderings. The corresponding lightest-state dip
sits at $E^{\rm NR}_{\rm l} \simeq (3$--$5)\times10^{5}$\,GeV, aligned
with the observed deficit at $E_{\rm dep} \sim 5\times10^{5}$\,GeV.
The best fit improves over the absence of $\nu$SI by
$\Delta\chi^2 \simeq 4.5$. This mild preference
leaves the no-$\nu$SI hypothesis still allowed.

As a result, the HESE data puts a lower bound on the
lightest mass for a mediator with $1 \lesssim \MP/{\rm MeV} \lesssim 4.7$
(NO) or $1 \lesssim \MP/{\rm MeV} \lesssim 4$ (IO).
This lower bound ranges from a few $10^{-3}$\,eV
near the best fit to a few $10^{-2}$\,eV
at the edge of the band. For instance, at
$\MP = 1$\,MeV we find $\ml > 7.76\times10^{-3}$\,eV (NO) and
$\ml > 7.82\times10^{-3}$\,eV (IO) at 95\% C.L.
Heavier mediators give no lower bound within the scanned range.

This lower bound actually does not rely on the choice $g_{\tau\tau} = 0.01$.
Scanning instead the ($\ml$, $g_{\tau\tau}$) plane at $\MP = 1$\,MeV, the
bound stays essentially unchanged for
$3\times10^{-4} \lesssim g_{\tau\tau} \lesssim 2\times10^{-2}$
since the coupling only affects the depth of the dips.

The bound is also stable against the neutrino and antineutrino
composition of the sources. Our main results assume $pp$ sources as
in the HESE analysis\,\cite{IceCube:2020wum} while $p\gamma$ sources
produce about twice as many neutrinos as antineutrinos. Since
neutrinos and antineutrinos have different deep-inelastic cross
sections\,\cite{Xie:2023suk}, the composition changes the predicted
event rates by a few percent. The resonance energy being
a function of only $\ml$ and $\MP$ are the same for both
sources. So the two bounds with different sources nearly
coincide although the $p \gamma$ case excludes a slightly
larger region, as shown in the Supplementary Material.

For comparison, we have also shown the DESI and big-bang
nucleosynthesis (BBN) bounds as shaded areas. The DESI bound
\cite{Kreisch:2019yzn,Berryman:2022hds} is derived within the
$\Lambda$CDM model and excludes the region above
2.3 (2.4) $\times 10^{-2}$\,eV for NO (IO).
At $\MP = 1$\,MeV
it combines with the HESE lower bound to confine the
lightest neutrino mass to
\begin{subequations}
\begin{align}
7.76 \times 10^{-3}\,{\rm eV} < m_1 &< 2.3\times10^{-2}\,{\rm eV}
\quad ({\rm NO}),
\\
7.82 \times 10^{-3}\,{\rm eV} < m_3 &< 2.4\times10^{-2}\,{\rm eV}
\quad ({\rm IO}).
\end{align}
\label{eq:window}
\end{subequations}
For NO and $1.92\,{\rm MeV} \leq \MP \leq 2.06\,{\rm MeV}$, the HESE
lower bound exceeds the DESI upper bound as shown by the tip stretching
into the DESI excluded region in \gfig{fig:Constraint} and no value of $\ml$ is
allowed. Besides, the light mediator below 1\,MeV cannot
satisfy the BBN constraint \cite{Blinov:2019gcj}.
Putting things together, a resonant $\nu$SI
in the MeV range would thus pin the lightest
neutrino mass to within a factor of a few around $10^{-2}$\,eV.

\vspace{2mm}
\textbf{Conclusion} --
The essential feature of an $s$-channel resonance
is that the center-of-mass energy is fixed to the
mediator mass,
$2 E_\nu (E_i - |\bm p| \cos\theta) \approx m^2_\phi$.
For the scattering with a low energy
C$\nu$B neutrino whose typical momentum is already
fixed by the Universe temperature, the resonance
energy $E_\nu$ of a UHE neutrino is then a function of the
tiny neutrino mass. Very differently from the two
non-relativistic neutrino mass eigenstates that
only allow narrow resonance
energy windows, the lightest neutrino can be
relativistic and induce a quite wide resonance
window at $10^6 \sim 10^8$\,GeV that can span
more than one order. In between, the shape
of the resonance energy window continuously
deforms with and leads to sensitivity on
the lightest neutrino mass.

In order to take into
consideration the different neutrino sources
and scattering cross section, it is necessary
to use the double transport equations
for both the neutrino and anti-neutrino modes.
In addition, the $s$-channel resonance has one
neutrino and one anti-neutrino in the final state.
It is then necessary to incorporate the double 
regeneration terms which can significantly
enhance the UHE neutrino flux below the resonance
windows.

Confronting with the HESE 7.5 years data observed
at IceCube clearly excludes the light-relic region
$1\,{\rm MeV} \lesssim \MP \lesssim 4.7$\,MeV for
couplings
$3\times10^{-4} \lesssim g_{\tau\tau} \lesssim 2\times10^{-2}$.
The combination with the DESI observation
confines the lightest mass to around $10^{-2}$\,eV
and some parameter space is fully excluded.

\vspace{2mm}
\textbf{Acknowledgements}
The authors are supported by the National Natural Science Foundation of China (12425506 and 12375101). JPP is also supported by the K.C. Wong Education Foundation.
SFG is also an affiliate member of Kavli IPMU, University of Tokyo.

\addcontentsline{toc}{section}{References}

\bibliographystyle{utphysGe}
\bibliography{ref}

\clearpage

\onecolumngrid

\clearpage

\setcounter{equation}{0}
\renewcommand{\theequation}{S\arabic{equation}}

\begin{center}
{\LARGE\bfseries Supplementary Material}
\end{center}

\vspace{1.cm} 

 \noindent
 {\large\bfseries 1. Double Transport Equations with Double Regeneration Terms}
 \vspace{0.5cm} 
 
We derive the double transport equations for the angularly
integrated comoving flux
$\phi_i \equiv a^3 |\bm p|^2 f_i(|\bm p|) / 2 \pi^2$ 
used in \geqn{eq:transport}. For ultra-high energy neutrinos,
their momenta are approximately their energies $E \simeq |\bm p|$. 
The transport equation of the comoving number density 
for the $i$-th neutrino mass eigenstate in the ultra
relativistic limit takes the form \cite{Ng:2014pca, Esteban:2021tub}
\begin{equation}
  \partial_t \phi_i
=
  {\cal S}_i + \partial_E (H E \phi_i)
- \Gamma_i^- \phi_i
+ \Gamma_i^+,
\end{equation}
where the subscript $i$ indicates the $i$-th mass
eigenstate.
The first term ${\cal S}_i$ is the astrophysical
UHE neutrino source defined in \geqn{eq:source}, 
$\Gamma_i^-$ is the loss rate for the $i$-th neutrino
due to the absorption process  and $\Gamma_i^+$
is the gain rate of the $i$-th neutrino from regenerations. 
While the source term increases the flux $\phi_i$,
the absorption and regeneration terms, $\Gamma^-_i$
and $\Gamma^+_i$, decrease and increase the flux,
respectively.

Converting the coordinate time $t$ to the redshift $z$ 
with $dt= -dz/H(1+z)$ yields 
\begin{equation}
  \partial_z \phi_i
=
- \frac {\mathcal S_i}{H (1+z)}
- \frac {\partial_E (E \phi_i)}{1+z}
+ \frac { \Gamma_i^- \phi_i 
        - \Gamma_i^+ }
        {H (1+z)},
\label{eq:sm_Boltzmann}
\end{equation}
where $H$ is the Hubble rate. Going back to the early
stage of our Universe, the coordinate time $t$
decreases while the redshift $z$ increases. This
opposite evolution trend explains the extra minus
sign.
\begin{figure}[h!]
\centering
\setlength{\unitlength}{1cm}
\thicklines

\par\medskip

\begin{minipage}{0.49\textwidth}
\centering
\small
\begin{picture}(6.6,2.2)(-0.8,0.1)
  \put(0.2,1.6){\line(2,-1){1.2}}
  \put(0.2,0.4){\line(2,1){1.2}}
  \put(3.0,1.0){\line(2,1){1.2}}
  \put(3.0,1.0){\line(2,-1){1.2}}

  \put(0.5,1.45){\vector(2,-1){0.3}}
  \put(1.1,0.85){\vector(-2,-1){0.3}}
  \put(3.3,1.15){\vector(2,1){0.3}}
  \put(3.9,0.55){\vector(-2,1){0.3}}

  \multiput(1.4,1.0)(0.16,0){10}{\line(1,0){0.09}}
  \put(1.4,1.0){\circle*{0.08}}
  \put(3.0,1.0){\circle*{0.08}}

  \put(0.1,1.72){\makebox(0,0)[r]{$\nu_k^{\mathrm{UHE}}$}}
  \put(0.1,0.28){\makebox(0,0)[r]{$\bar\nu_l^{\mathrm{C}\nu\mathrm{B}}$}}
  \put(2.2,1.25){\makebox(0,0){$\phi$}}
  \put(4.25,1.72){\makebox(0,0)[l]{$\nu_i$}}
  \put(4.25,0.28){\makebox(0,0)[l]{$\bar\nu_j$}}
  \put(4.8,1.45){\makebox(0,0)[l]{$\Gamma_i^{\nu\nu}$}}
  \put(4.8,0.55){\makebox(0,0)[l]{$\Gamma_j^{\nu\bar\nu}$}}
\end{picture}
\end{minipage}
\hspace{-1cm}
\begin{minipage}{0.49\textwidth}
\centering
\small
\begin{picture}(6.6,2.2)(-0.8,0.1)
  \put(0.2,1.6){\line(2,-1){1.2}}
  \put(0.2,0.4){\line(2,1){1.2}}
  \put(3.0,1.0){\line(2,1){1.2}}
  \put(3.0,1.0){\line(2,-1){1.2}}

  \put(1.1,1.15){\vector(-2,1){0.3}}
  \put(0.5,0.55){\vector(2,1){0.3}}
  \put(3.3,1.15){\vector(2,1){0.3}}
  \put(3.9,0.55){\vector(-2,1){0.3}}

  \multiput(1.4,1.0)(0.16,0){10}{\line(1,0){0.09}}
  \put(1.4,1.0){\circle*{0.08}}
  \put(3.0,1.0){\circle*{0.08}}

  \put(0.1,1.72){\makebox(0,0)[r]{$\bar\nu_k^{\mathrm{UHE}}$}}
  \put(0.1,0.28){\makebox(0,0)[r]{$\nu_l^{\mathrm{C}\nu\mathrm{B}}$}}
  \put(2.2,1.25){\makebox(0,0){$\phi$}}
  \put(4.25,1.72){\makebox(0,0)[l]{$\nu_i$}}
  \put(4.25,0.28){\makebox(0,0)[l]{$\bar\nu_j$}}
  \put(4.8,1.45){\makebox(0,0)[l]{$\Gamma_i^{\bar\nu\nu}$}}
  \put(4.8,0.55){\makebox(0,0)[l]{$\Gamma_j^{\bar\nu\bar\nu}$}}
\end{picture}
\end{minipage}

\caption{The resonant $s$-channel scattering of
an UHE neutrino (left) or anti-neutrino (right) with
thermal C$\nu$B mediated by a scalar $\phi$.
For regeneration, the four terms $\Gamma^{\alpha \beta}$
are labeled by the incident species $\alpha$ and the regenerated
species $\beta$ while their subscripts denote the regenerated
mass eigenstates.
}
\label{fig:Feynman}
\end{figure}
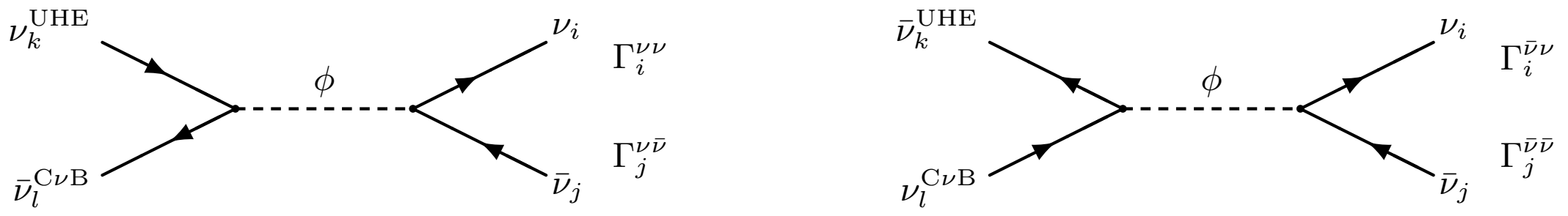

For the UHE neutrinos scattering with C$\nu$B mediated
by $\nu$SI, the incoming UHE neutrinos could be either
neutrinos or antineutrinos. Note that they have different
fractions from the astrophysical sources. For example,
the $p \gamma$ process tends to produce a positively
charged $\pi^+$ and hence one anti-neutrino $\bar \nu_\mu$
besides two neutrinos $\nu_e$ and $\nu_\mu$ in the
final state. On the other hand, the $pp$ source produces
a pair of charged pions $\pi^+ \pi^-$ which subsequently
leads to equal amount of neutrinos and anti-neutrinos.
To take all possible sources into account, it is necessary
to consider double transport equations for neutrinos
and anti-neutrinos,
\begin{subequations}
\begin{align}
  \partial_z \phi_i^\nu
& =
- \frac {\mathcal S_i^{\nu}}{H (1+z)}
- \frac {\partial_E (E \phi_i^{\nu})}{1+z}
+ \frac { \Gamma_i^\nu \phi_i^\nu 
         - \Gamma^{\nu \nu}_i
         - \Gamma^{\bar \nu \nu}_i}
        {H (1+z)},
\label{Supp:eq:transport_nu} 
\\
  \partial_z \phi_i^{\bar \nu}
& =
- \frac {\mathcal S_i^{\bar \nu}}{H (1+z)}
- \frac {\partial_E (E \phi_i^{\bar \nu})}{1+z}
+ \frac { \Gamma_i^{\bar \nu} \phi_i^{\bar \nu} 
         - \Gamma^{\bar \nu \bar \nu}_i
         - \Gamma^{\nu \bar \nu}_i}
        {H (1+z)},
\label{Supp:eq:transport_antinu} 
\end{align}
\label{Supp:eq:transport} 
\end{subequations}
which is exactly a copy of \geqn{eq:transport}
that we show here for the completeness of this
supplementary material.

Besides absorbing the initial-state neutrino
$\nu^{\rm UHE}_k$ or anti-neutrino $\bar \nu^{\rm UHE}_k$,
the same process in \gfig{fig:Feynman} can
regenerate both neutrinos and anti-neutrinos in
the final states. With a single particle being
absorbed, two particles can be regenerated. So
the scattering process can really enhance the
flux. Another difference from the transport
equation adopted in the literature \cite{Esteban:2021tub}
is that the double regeneration terms,
$\Gamma^{\nu \nu}_i$ and $\Gamma^{\bar \nu \nu}_i$
for the neutrino mode and
$\Gamma^{\bar \nu \bar \nu}_i$ and $\Gamma^{\nu \bar \nu}_i$
for the anti-neutrino mode, are used in
\geqn{Supp:eq:transport}. The propagation of
UHE neutrinos should be described by the double
transport equations with double regeneration
terms. 

Below we will first elaborate the absorption and
regeneration terms before discussing the effect
of different astrophysical sources of UHE neutrinos.

\vspace{1cm}
\noindent
{\large\bfseries 2. Absorption and Regeneration Terms}

\vspace{0.5cm}

Note that the scalar $s$-channel scattering has
an isotropic decay distribution and equal charge-conjugate
branching probabilities to give equal neutrino 
and antineutrino daughter kernels in the
ultrarelativistic limit,
$\Gamma_i^{\nu \nu} = \Gamma_i^{\nu \bar \nu}$ and
$\Gamma_i^{\bar \nu \nu} = \Gamma_i^{\bar \nu \bar \nu}$.
For a charge-symmetric relic background, all four
regeneration terms coincide when the incident
neutrino and antineutrino spectra are equal. 

In principle, a scalar mediator can also induce
$t$-channel and $u$-channel diagrams for the
two neutrino scattering process. However, these
two extra contributions are suppressed than their
$s$-channel counterpart in \gfig{fig:Feynman}.
With resonance, the $s$-channel cross section is
proportional to $g^2$ where $g$ denotes the coupling
strength. However, the $t$- and $u$-channels is
proportional to $g^4$ with extra suppression of
$g^2$. It is safe to omit the $t$- and $u$-channels
in the current study.

\vspace{2mm}
\textbf{The absorption rate} --
A UHE neutrino $\nu_i(p)$ is removed from the spectrum by the resonant
scattering $\nu_i(p) + \bar{\nu}_j(k) \to \nu_k(p') + \bar{\nu}_l(k')$ 
on a relic $\nu_j(k)$. Its rate is
\begin{align}
\Gamma^{\nu/\bar{\nu}}_{i}(z,E)
\equiv \ &
\frac{1}{2E}
\int \frac{d^3\boldsymbol{k}}{(2\pi)^3 2E_k}
\frac{d^3 \boldsymbol{p}'}{(2\pi)^3 2E_{p'}}
\frac{d^3\boldsymbol{k}'}{(2\pi)^3 2E_{k'}}
(2\pi)^4 \delta^{(4)}(p + k - p' - k')
\nonumber\\
& \times \sum_{j,k,l} |\mathcal{M}_s|^2_{ij \to kl}\,
f_{\rm C\nu B}^j(|\bm k|)
\left[1-f^k(|\bm p'|)\right]\left[1-f^l(|\bm k'|)\right],
\label{eq:abs_def}
\end{align}
where only the resonant $s$-channel amplitude $\mathcal{M}_s$ is kept.
$f^j(|\bm k|)$ denotes the distribution function of the j-mass eigenstate
neutrino with momentum $\bm k$.
The final-state neutrinos are ultra-relativistic and their occupation
numbers are far below unity, so we set $1-f \simeq 1$. The integral
over the final states then defines the $2\to2$ cross section,
\begin{equation}
\sigma_{ij \to kl}(s)
\equiv
\frac{1}{4F}
\int \frac{d^3p'}{(2\pi)^3 2E_{p'}}
\frac{d^3k'}{(2\pi)^3 2E_{k'}}
(2\pi)^4 \delta^{(4)}(p + k - p' - k')\,
|\mathcal{M}_s|^2_{ij \to kl},
\label{eq:2_to_2_xsection}
\end{equation}
with the flux factor $F \equiv p \cdot k = E E_k v_{\rm rel}$ and the
M{\o}ller velocity $v_{\rm rel} = s/(2EE_k) = 1 - |\bm k|\cos\theta/E_k$,
where $\theta$ is the angle between the two initial momenta. This gives
\geqn{eq:Gamma-} of the main text,
\begin{equation}
\Gamma^{\nu/\bar{\nu}}_{i}(z,E)
=
\sum_{j,k,l}
\int \frac{d^3 \bm k}{(2\pi)^3}\, f_{\rm C\nu B}^j(|\bm k|)\,
v_{\rm rel}  \sigma_{ij \to kl}(s).
\end{equation}

The dependence on the relic kinematics enters through the angular
integration. Since the cross section depends on the angle only through
$s = 2E(E_k - |\bm k|\cos\theta)$, we 
switch the variable $\cos\theta$ to $s$
with
$ds = -2E|\bm k|\,d\cos\theta$,
\begin{equation}
\Gamma^{\nu/\bar{\nu}}_{i}(z,E)
=
\sum_{j}
\frac{1}{4\pi^2}
\int_0^\infty d|\bm k|\,|\bm k|^2 f_{\rm C\nu B}^j(|\bm k|)
\int_{s_-}^{s_+} \frac{ds}{2E|\bm p|}\,
v_{\rm rel} \sum_{kl} \sigma_{ij \to kl}(s),
\qquad
s_\pm \equiv 2E\left(E_k \pm |\bm p|\right).
\label{eq:Gamma_s}
\end{equation}
The Breit--Wigner distribution of \geqn{eq:sigma_BW} is sharply peaked at
$s = \MP^2$, so the M{\o}ller velocity can be evaluated there,
$v_{\rm rel} \simeq \MP^2/(2EE_k)$. The remaining integral is analytic,
\begin{equation}
\int_{s_-}^{s_+} ds\, \sigma_{ij \to kl}(s)
=
\frac{|g_{ij}|^2 |g_{kl}|^2}{16\pi}
\left[\mathcal{F}(s_+) - \mathcal{F}(s_-)\right],
\quad
\mathcal{F}(s)
\equiv
\frac{1}{2} \ln\!\left[(s-\MP^2)^2 + \MP^2 \Gamma_\phi^2\right]
+
\frac{\MP}{\Gamma_\phi}
\arctan\!\left(\frac{s-\MP^2}{\MP \Gamma_\phi}\right).
\label{eq:int_cs}
\end{equation}
For $\Gamma_\phi \ll \MP$, the arctangent reduces to a step function
and the logarithmic term is negligible,
$\mathcal{F}(s_+) - \mathcal{F}(s_-) \simeq (\pi\MP/\Gamma_\phi)\,
\Theta(s_+ - \MP^2)\,\Theta(\MP^2 - s_-)$. With
$\Gamma_\phi = \MP\sum_{kl}|g_{kl}|^2/16\pi$, the sum over final states
cancels against the width, and
\begin{equation}
\Gamma^{\nu/\bar{\nu}}_{i}(z,E)
\simeq
\sum_{j,k,l}
\frac{|g_{ij}|^2 \MP^2}{16\pi E^2}
\int_0^\infty d|\bm k|\,
\frac{|\bm k|\, f_{\rm C\nu B}^j(|\bm k|)}{E_k}\,
\Theta(s_+ - \MP^2)\,\Theta(\MP^2 - s_-)
.
\label{eq:Gamma_narrow}
\end{equation}
The resonant window is determined by the condition $s_- \le \MP^2 \le s_+$, with $s_{\pm}$ corresponding to $E_j^{\pm}$ of \geqn{eq:res_window}. 
Absorption is thus the resonance window of the
main text averaged over the thermal relic distribution. It depends on
the coupling only through $|g_{ij}|^2$ and is independent of the
natural width.

\vspace{2mm}
\textbf{The regeneration term} --
The same process, $\nu_k(p') + \nu_l(k') \to \nu_i(p) + \nu_j(k)$,
regenerates two UHE neutrinos, $\nu_i(E)$ and $\bar{\nu}_j(E_k)$,
from UHE neutrinos $\nu_k$ with higher
energy $E' \ge E,E_k$ scattering on relics $\nu_l$. The corresponding 
regeneration term for the outgoing neutrino $\nu_i(p)$ from the 
incoming neutrino $\nu_k(p')$ is 
\begin{equation}
\Gamma^{\nu\nu}_i(z,E)
\equiv
\frac{a^3|\bm p|^2}{2\pi^2}
\sum_{k,l,j}
\frac{1}{2E}
\int \frac{f_{\rm C\nu B}^l(|\bm k'|) d^3 \bm k'}{(2\pi)^3 2E_{k'}}
\frac{f^k(|\bm p'|) d^3 \bm p'}{(2\pi)^3 2E_{p'}}
\frac{d^3 \bm k}{(2\pi)^3 2E_k}
(2\pi)^4 \delta^{(4)}(p' + k' - p - k)\,
|\mathcal{M}_s|^2_{kl \to ij},
\label{eq:regeneration_diluted}
\end{equation}
where again $1 - f^j \simeq 1$ for the final-state
relic neutrinos. The
integral over $\bm k$ is related to the differential cross section in
the energy $E$ of the regenerated neutrino. Integrating over its
direction, one obtains
\begin{equation}
\int \frac{d^3\bm k}{(2\pi)^3 2E_k}
(2\pi)^4 \delta^{(4)}(p' + k' - p - k)\,
|\mathcal{M}_s|^2_{kl \to ij}
=
\frac{d^2 \sigma^{\nu \nu}_{kl \to ij}}{d|\bm p| d\Omega_p}\,
\frac{8F'(2\pi)^3}{E},
\qquad
F' \equiv E_{k'} E_{p'} v'_{\rm rel},
\end{equation}
where $\sigma^{\nu \nu}$ denotes the fact that
we are considering the scattering process of
an initial-state UHE neutrino into a final-state
UHE neutrino with $E = p^0$. Inserting this into
\geqn{eq:regeneration_diluted} and writing
$d^3\bm p' = E_{p'}^2 dE_{p'} d\Omega'$, we get
\begin{align}
    \Gamma^{\nu\nu}_{i}(z,E)
=
    \sum_{k,l,j}
    \int 
    \phi_k^{\nu}(|\bm p'|) d |\bm p'|
    \int \frac{f_{\rm C\nu B}^{l}(|\bm k'|)d^3\bm k'}{(2\pi)^3}
    v_{\rm rel}' \frac{d \sigma^{\nu \nu}_{kl \to ij}}{d|\bm p|},
\end{align}
where we use the definition of comoving
diffused flux
$\phi \equiv a^3 v \frac{|\bm p|^2}{2\pi^2} f (|\bm p|)$
which is obtained by integrating out the solid angle $d\Omega'$.

Similarly, the regeneration term for the outgoing neutrino 
$\nu_{i}(p)$ from the incoming anti-neutrino $\bar{\nu}_k(p^{\prime})$
is 
\begin{align}
    \Gamma^{\bar{\nu}\nu}_{i}(z,E)
=
    \sum_{k,l,j}
    \int 
    \phi_k^{\bar{\nu}}(|\bm p'|) d |\bm p'|
    \int \frac{f_{\rm C\nu B}^{l}(|\bm k'|)d^3\bm k'}{(2\pi)^3}
    v_{\rm rel}' \frac{d \sigma^{\bar \nu \nu}_{kl \to ij}}{d|\bm p|}.
\end{align}
Therefore, the regeneration term which describes the scattering 
process with the incoming neutrino species $a$ and 
outgoing neutrino species $b$ can be expressed as a general 
form $\Gamma^{ab}_{i}$, 
\begin{align}
    \Gamma^{ab}_{i}
\equiv 
     \sum_{k,l,j}
    \int 
    \phi_k^{a}(|\bm p'|) d |\bm p'|
    \int \frac{f_{\rm C\nu B}^{l}(|\bm k'|)d^3\bm k'}{(2\pi)^3}
    v_{\rm rel}' \frac{d \sigma^{ab}_{kl \to ij}}{d|\bm p|}
\end{align}
where $a,b \in \{\nu,\bar{\nu}\}$.
For a scalar $s$-channel exchange, the final state is isotropic in the
center-of-mass frame, $d\sigma/d\cos\theta^* = \sigma/2$. Boosting to
the cosmic frame, $E = \gamma(E^* + \beta |\bm p^*|\cos\theta^*)$ with
$|\bm p^*| = \sqrt{s}/2$ and $\gamma\beta = |\bm P|/\sqrt{s}$, where
$\bm P = \bm p' + \bm k'$, gives a flat spectrum,
\begin{equation}
  \frac{d \sigma^{ab}_{kl \to ij}}{d|\bm p|}
=
  \frac{\sigma^{ab}_{kl \to ij}(s)}{|\bm P|},
\qquad 0 \lesssim E \le E_{p'}.
\end{equation}
The angular integral is performed as for the absorption rate, changing $\cos\theta'$ to $s$ 
with $s_\pm' = 2E_{p'}(E_{k'} \pm |\bm k'|)$ and
evaluating $v'_{\rm rel}/|\bm P| \simeq \MP^2/(2E_{p'}^2E_{k'})$ at the
resonance,
\begin{equation}
    \Gamma^{\nu\nu}_i(z,E)
=
    \sum_{k,l,j}
    \int \phi_{k}^{\nu}(|\bm p'|) d|\bm p'|
    \frac{1}{4\pi^2}
    \int_0^{\infty} d|\bm k'| |\bm k'|^2 f^l_{\rm C\nu B}(|\bm k'|)
    \frac{1}{2E_{p'}|\bm k'|} \frac{m_\phi^2}{2E_{p'}^2E_{k'}}
    \frac{|g_{kl}|^2|g_{ij}|^2}{16\pi}
    \left[\mathcal{F}(s'_+) - \mathcal{F}(s'_-)\right]
\end{equation}
We solve \geqn{eq:transport_nu} and
\geqn{eq:transport_antinu} with these rates by
generalizing the package $\mathtt{nuSIProp}$
\cite{Esteban:2021tub}, in which the absorption
and regeneration term are replaced by the thermal
averages above.

\vspace{1cm}

\noindent
{\large\bfseries 3. Comparison between Sources}

\vspace{0.5cm} 

The ultra-high-energy astrophysical neutrinos 
can originate from different production mechanisms 
with different astrophysical sources, 
leading to distinct neutrinos and anti-neutrinos 
compositions. There are two usual and representative 
scenarios are hadronuclear ($pp$) and photohadronic ($p\gamma$)
sources. 
In the $pp$ scenarios, which is expected to 
be dominant in star-forming galaxies\,\cite{Skrzypek:2025tmg}, 
the amounts of neutrinos are approximately equal 
to that of anti-neutrinos. 
However, in the $p\gamma$ process, the amounts 
of neutrinos are nearly twice as large as
that of anti-neutrinos\,\cite{Skrzypek:2025tmg,Liu:2023lxz}. 
So the astrophysical sources will affect the initial
neutrino flux compositions.

\begin{figure}
\centering
\includegraphics[width=0.65\columnwidth]{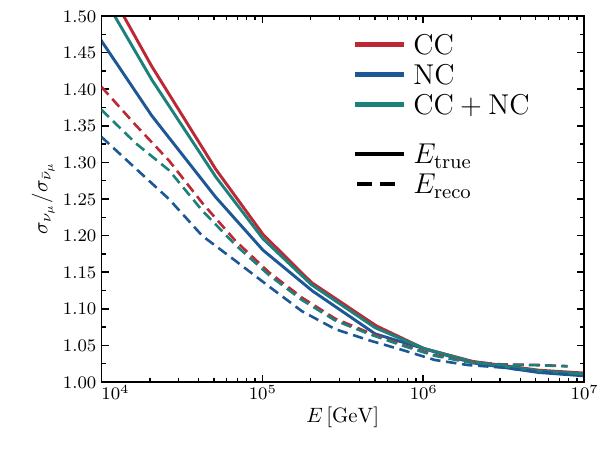}
\caption{The ratio between the $\nu_\mu$ to $\bar \nu_\mu$
deep-inelastic scattering (DIS) cross sections. 
The red line denotes the charged-current (CC) interaction, 
the blue line denotes the neutral-current (NC) interaction
and the green line represents the combined (CC + NC) interactions.
Solid curves show the 
cross section ratios with the true neutrino energy \cite{Xie:2023suk}, 
while dashed 
curves denote the corresponding effective ratios with the 
reconstructed deposit energy.
}
\label{fig:DIS_xsec}
\end{figure}

This difference is relevant for IceCube observations
because neutrinos and anti-neutrinos have different 
deep-inelastic scattering (DIS) cross sections 
with nucleons, due to the chiral structure 
of weak interactions and the different 
quark and anti-quark distributions inside nucleons\,\cite{Xie:2023suk,Ge:2013ffa}.
The difference is energy-dependent, as illustrated in \gfig{fig:DIS_xsec},
where we show the ratio between $\nu_{\mu}$ and $\bar{\nu}_\mu$
cross section for CC, NC and the total contribution\,\cite{Xie:2023suk}.
The difference can reach 20\% around $10^5$\,GeV
and could further enhance to 50\% around $10^4$\,GeV.
With the energy transfer matrix provided by HESE-7.5 years analysis\,\cite{IceCube:2020wum}, 
we also get the ratio with the reconstructed 
energy. Although the difference is reduced 
after accounting for the detector's energy reconstruction, 
the neutrino and antineutrino scattering cross sections 
remain different, particularly at energies below $10^6\, \mathrm{GeV}$. The reduction from the
true energy case to the reconstructed one is
typically at the level of 40\%.

\begin{figure}[t]
\centering
\includegraphics[width=0.48\textwidth]{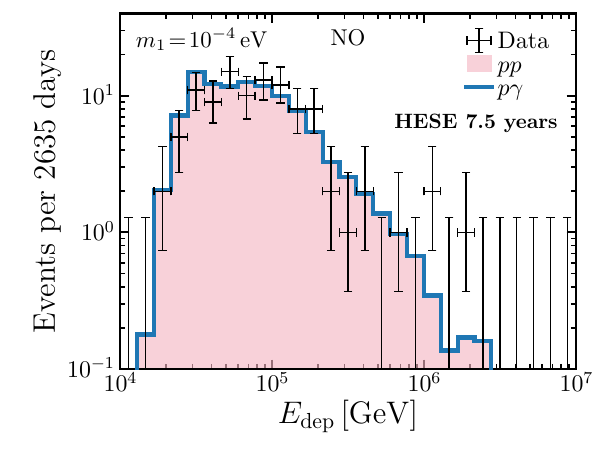}
\hfill
\includegraphics[width=0.48\textwidth]{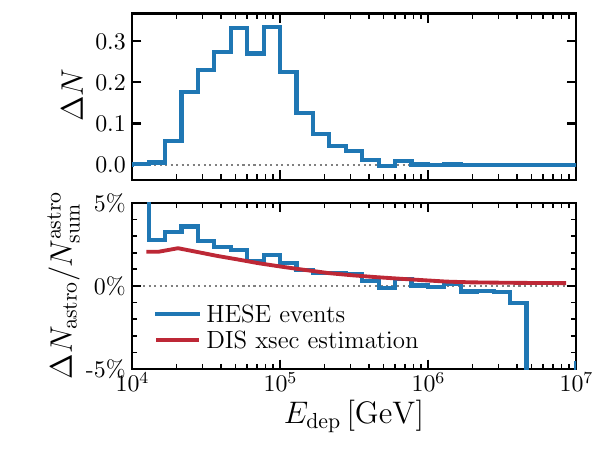}
\caption{(Left) The predicted HESE event distributions
using the $pp$ (pink shaded region) and $p\gamma$ 
(blue line) source scenarios as a function of
reconstructed deposit energy $E_{\rm dep}$, 
with $m_1 = 10^{-4}\,\rm eV$ and normal ordering (NO). 
The black points with error bar shows the observed data.
(Right)
The absolute difference between the predicted event
numbers $\Delta N = N_{p\gamma} - N_{pp}$ (upper
right) and the relative difference (lower right).
While the blue line represents the actually
simulated event numbers, the red curve shows
the estimation with the neutrino and anti-neutrino
DIS cross section differences.
}
\label{fig:Nu_Events_compare}
\end{figure}

Therefore, both the
differences of the neutrino flux
compositions and DIS cross sections will affect our 
predicted event numbers and then have the impacts
on the final constraints. \gfig{fig:Nu_Events_compare}
shows the difference 
on the event numbers, with the same benchmark choice
as \gfig{fig:Event_Number}. The left panel shows the event numbers for 
$pp$ and $p\gamma$ sources. To see the difference,
the upper right panel shows the absolute difference 
$\Delta N = N_{p\gamma} - N_{pp}$
in the predicted HESE event numbers 
between the \(p\gamma\) and \(pp\) scenarios.
The difference is most pronounced at lower 
deposit energies, from $2 \times 10^{4} \,\rm GeV$ 
to $4 \times 10^{5} \, \rm GeV$. Also, this
difference gradually decreases as the energy
increases. To quantify this effect, 
the lower right panel shows
the relative difference, $\Delta N / (N_{p\gamma} + N_{pp})$.
The difference is typically at the level of a few percent 
and becomes smaller at higher energies.

To fully understand the results, we put a red
curve to show a simplified estimate based on 
the neutrino and anti-neutrino DIS cross sections 
and compositions. The event number is proportional to the 
flux and the cross section, $N \propto \phi \times \sigma(E)$.
Assume that the total neutrino flux is the same for $pp$ and $p\gamma$
source scenarios, denoted by $\Phi$. Then for $pp$ source scenario, 
neutrino flux $\phi_\nu^{pp} \propto \frac{1}{2} \Phi$ and 
also anti-neutrino flux $\phi_{\bar{\nu}}^{pp} \propto \frac{1}{2}\Phi$.
For the $p\gamma$ source scenario, neutrino flux 
$\phi_\nu^{p\gamma} \propto \frac{2}{3} \Phi$, with anti-neutrino flux
$\phi_{\bar{\nu}}^{p\gamma} \propto \frac{1}{3} \Phi$. 
Then we can get the event numbers for these two source scenarios,
\begin{align}
  N_{pp}
\propto
  \frac {\sigma_\nu + \sigma_{\bar{\nu}}} 2
  \Phi,
\qquad
  N_{p\gamma}
\propto
  \frac {2\sigma_\nu + \sigma_{\bar{\nu}}} 3
  \Phi.
\end{align}
Then the relative difference for two source scenarios is estimated by
\begin{align}
    \frac{\Delta N}{N_{p\gamma}+N_{pp}}
&=
    \frac{
    \left(\frac{2}{3}\Phi\sigma_{\nu}
    +\frac{1}{3}\Phi\sigma_{\bar{\nu}}\right)
    -
    \left(\frac{1}{2}\Phi\sigma_{\nu}
    +\frac{1}{2}\Phi\sigma_{\bar{\nu}}\right)
    }{
    \left(\frac{2}{3}\Phi\sigma_{\nu}
    +\frac{1}{3}\Phi\sigma_{\bar{\nu}}\right)
    +
    \left(\frac{1}{2}\Phi\sigma_{\nu}
    +\frac{1}{2}\Phi\sigma_{\bar{\nu}}\right)
    }
=
    \frac{\sigma_{\nu}-\sigma_{\bar{\nu}}}
    {7\sigma_{\nu}+5\sigma_{\bar{\nu}}}.
\end{align}
For example, at the deposited energy $E_{\rm dep} =10^5\,\rm GeV$, 
the neutrino-to-antineutrino DIS cross section ratio is about $1.15$. 
However, the relative difference in event number
is about $1\%$.

\begin{figure}[b]
\centering
\includegraphics[width=0.68\columnwidth]{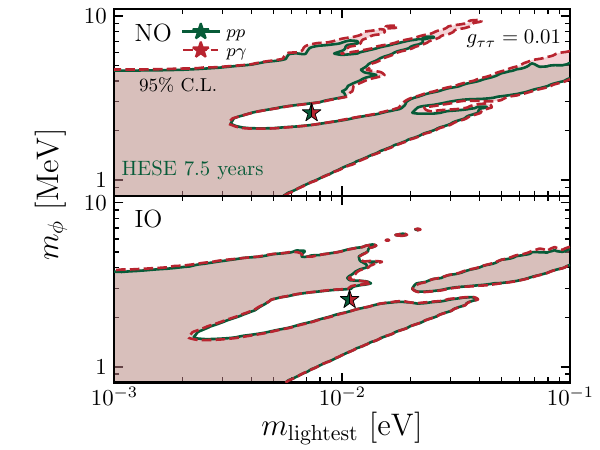}
\caption{Comparison of the $95\%$\,C.L. exclusion regions on the 
($m_{\rm lightest}, m_\phi$) plane by the HESE 7.5-year data
for $pp$ (green solid)
and $p\gamma$ (red dashed) sources with
stars marking the corresponding best-fit values.
The upper and lower panels are for the normal
and inverted mass orderings, respectively.}
\label{fig:Nu_Constraint_Compare}
\end{figure}
Finally, we compare the constraints on the lightest neutrino
mass obtained from $pp$ source and $p\gamma$ source scenarios, 
as shown in \gfig{fig:Nu_Constraint_Compare}.
Despite the different neutrino and anti-neutrino
ratios, the resulting exclusion boundaries nearly
overlap with only minor differences in 
some regions of the parameter space. 
This indicates that the constraints 
on the lightest neutrino mass are not that much
sensitive to the choice between these two benchmark sources.
The reason is that the resonance kinematics is determined by
the lightest neutrino mass and the mediator mass, which is 
independent of the neutrino sources.

\end{document}